\documentclass[aps,prl,cha,graphicx,amsmath,amssymb,reprint,twocolumn,
superscriptaddress]{revtex4}
\usepackage{graphicx}
\newcommand{\be}{\begin{equation}}
\newcommand{\ee}{\end{equation}}
\newcommand{\bea}{\begin{eqnarray}}
\newcommand{\eea}{\end{eqnarray}}
\newcommand{\nn}{\nonumber}
\newcommand{\bsf}[1]{\textsf{\textbf{#1}}}

\begin{document}
\title{The Phase Synchronized State of Oriented Active Fluids}
\author{Sebastian F\"urthauer}
\affiliation{TIFR Centre for Interdisciplinary Sciences, Tata Institute of Fundamental
Research, 21 Brundavan Colony, Narsingi, Hyderabad 500 075, India}
\email{sebastian.fuerthauer@gmail.com,sriram@tifrh.res.in}
\author{Sriram Ramaswamy}
\thanks{on leave from Dept of Physics, Indian Institute of Science, Bangalore}
\affiliation{TIFR Centre for Interdisciplinary Sciences, Tata Institute of Fundamental
Research, 21 Brundavan Colony, Narsingi, Hyderabad 500 075, India}

\date{\today}
\begin{abstract}
We present a theory for self-driven fluids, such as
motorized cytoskeletal extracts or microbial suspensions, that takes into
account the underlying periodic duty cycle carried by the constituent active
particles. We show that an orientationally ordered
active fluid can undergo a transition to a state in which the
particles synchronize their phases. This spontaneous breaking of
time-translation invariance gives rise to flow instabilities distinct from those
arising in phase-incoherent active matter.
Our work is of relevance to the transport of fluids in living systems, and
makes predictions for concentrated active-particle suspensions and optically
driven colloidal arrays.
\end{abstract}
\maketitle

\begin{figure}
 \includegraphics[width=\columnwidth]{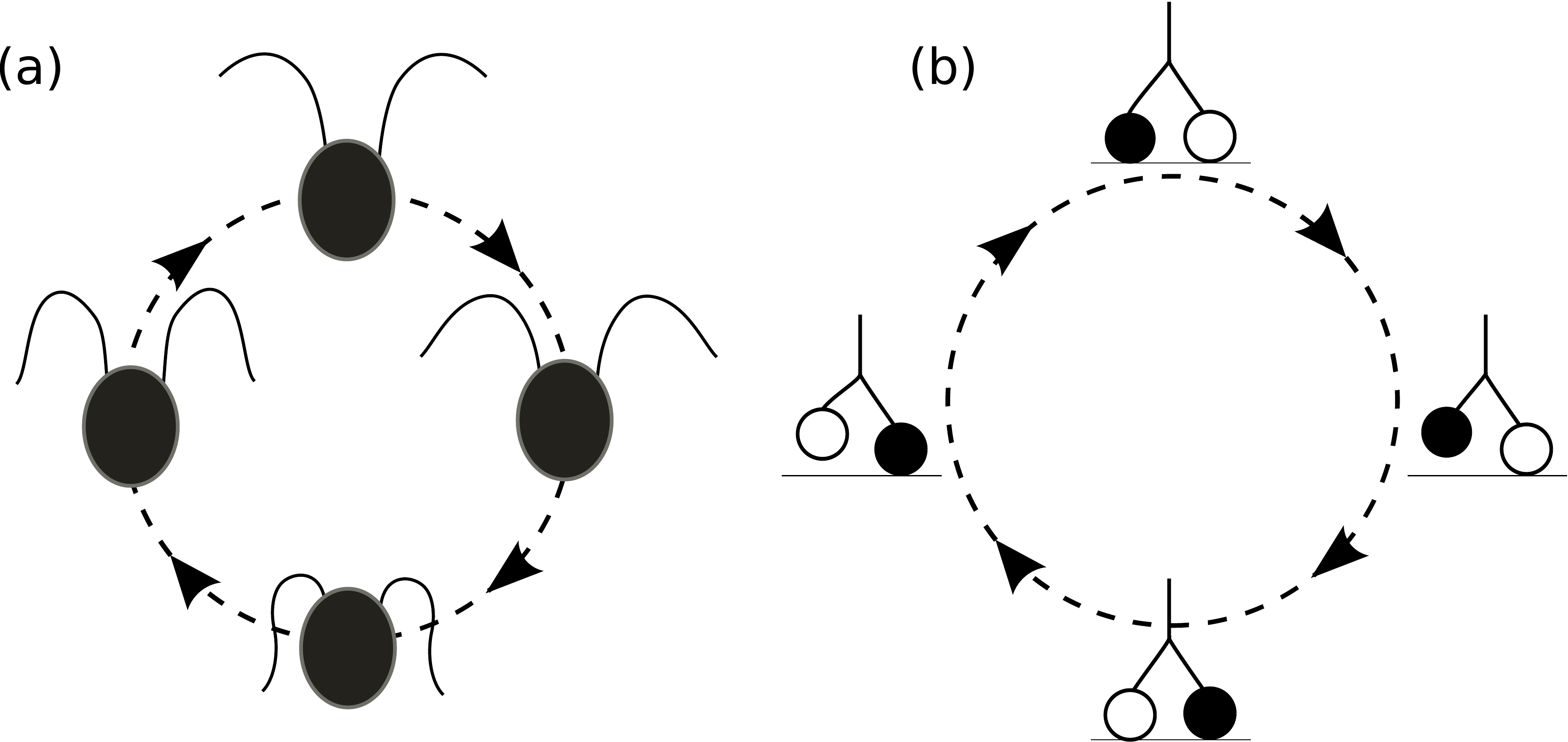} 
 \caption{Duty cycles of active processes. Sketch of (a) the swimming stroke of the alga {\it chlamydomonas} and 
 (b) a molecular motor moving along a cytoskeletal filament.}
 \label{fig:fig1}  
 \end{figure}

%
The field of broken-symmetry hydrodynamics \cite{mart72} is seeing increasing success in
describing the large-scale behavior of spatially ordered active systems
\cite{toner, jul07, rama10, marc13, ezhi13}. In this Letter we explore the
consequences of the spontaneous breaking of \textit{time}-translation
invariance \cite{stro04,stro05} for orientationally ordered fluids, uncovering 
the existence of a phase coherent state of active matter, which 
we expect to be seen in experiments and to be relevant to fluid transport 
in biological systems \cite{guir07}. 

Time-periodicity in a macroscopic system can arise either through
large-scale spontaneous wavelike motion in a system without
distinct local oscillators \cite{salb07,giom08} or via phase-locking of
particle-scale periodic processes \cite{stro04,stro05,uchi1011,brum12}, the case
with which this Letter is concerned. Active particles \cite{schw07}, such as
motor proteins \cite{howa01, chow13}, motile organisms \cite{bray00}, or
non-living
imitations thereof \cite{paxt04,hows07,nara07,webe13} convert energy from a reservoir into mechanical work.
This energy conversion process generally involves a periodic duty
cycle (Fig.~\ref{fig:fig1}), quantified by a phase variable
$\theta\in[0,2\pi]$ and is responsible for the force- and
torque-free motion of an active particle. Hydrodynamic theories
for fluid or liquid-crystalline phases of active particles -- active fluids, for
short -- have been developed and predict that these materials can flow
spontaneously if the stresses generated by the active particles are large enough
\cite{simh02,live03,hatw04,krus0405,voit05,mare07,joan07,soko09,fuer12tc}.
Many biological materials are active fluids
\cite{maye10,krus13,need13}.
The emergence of local oscillators in active systems was treated in \cite{juli97,lask13}.
 It is natural to ask how synchronization physics
\cite{stro04,stro05,uchi1011} enriches active-matter
hydrodynamics, usually presented in terms of the force and torque dipoles that
emerge from an incoherent time-average of the underlying duty cycles.
In this Letter we explore the broken-symmetry hydrodynamics of a
phase-coherent active fluid \cite{footnote_leon}.

\begin{figure}[h]
 \includegraphics[width=\columnwidth]{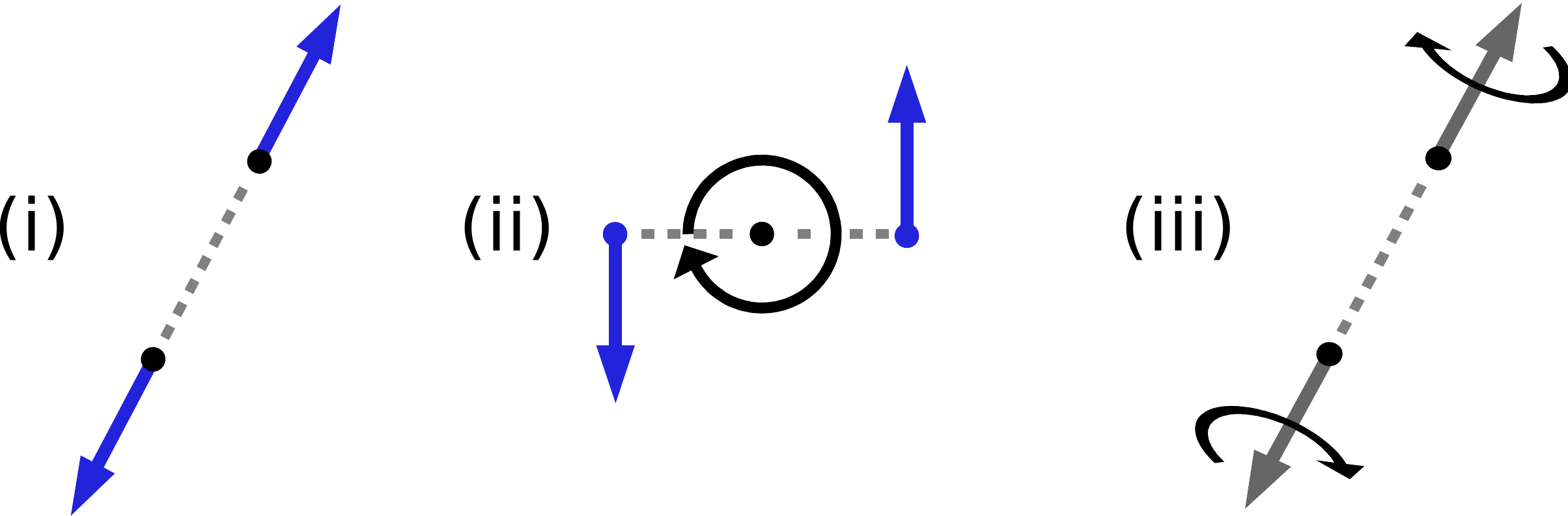} 
 \caption{Sketches of active shakers (i), polar rotors (ii) and nematic rotors
(iii). These objects exert periodic forces (blue arrows) and torques (black
curved arrows) on the fluid. Note that the total torque and force exerted by
(i), (ii) and (iii) is zero at each instant in time.}
\label{fig:fig2}  
\end{figure}

\begin{figure}
 \includegraphics[width=\columnwidth]{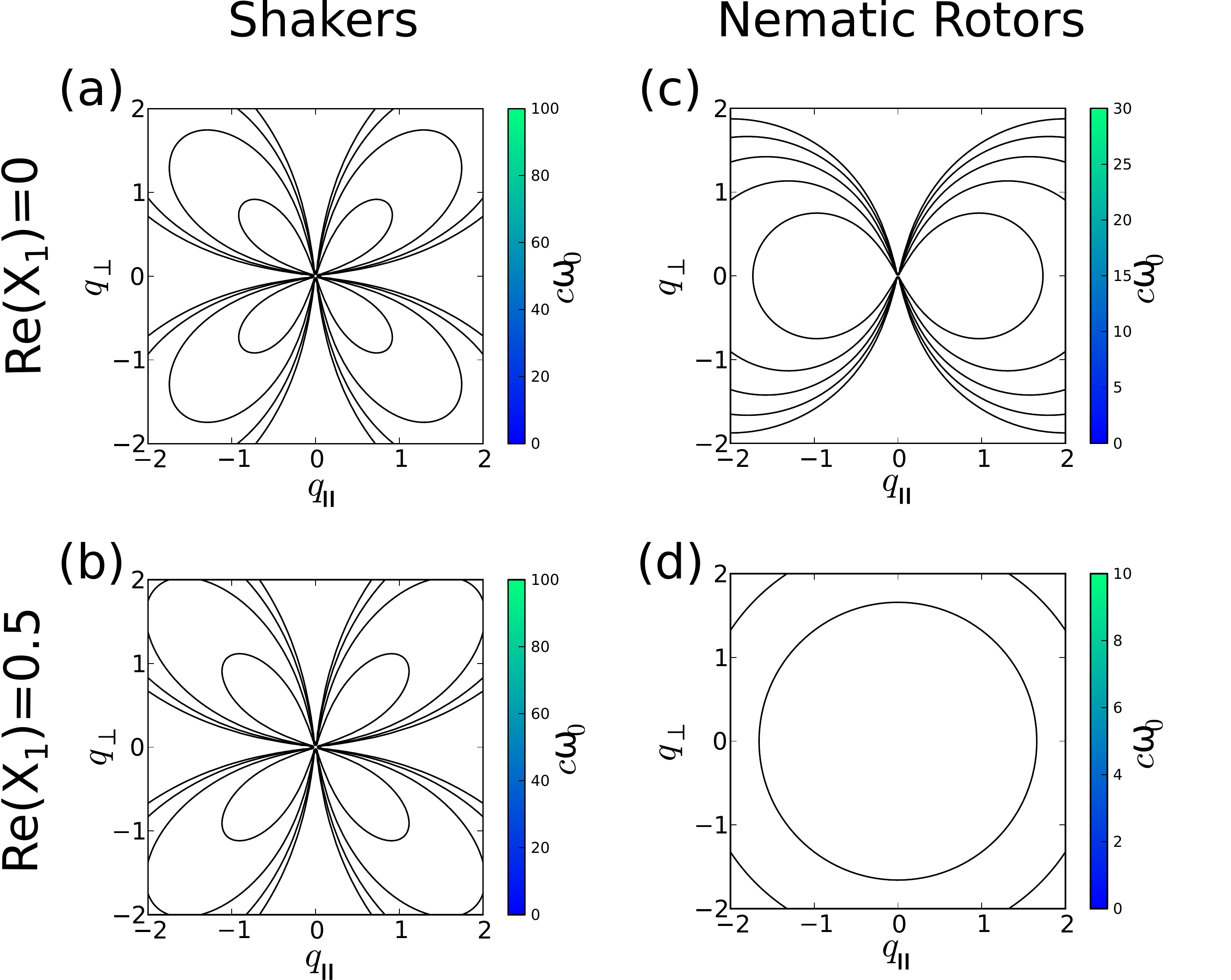} 
 \caption{Critical frequency for the onset of synchronization for
 shakers(a,c) and nematic rotors (b,d). We show $X_1=-i$ (a,b) and $X_1=0.5-i$ (c,d). Black contour lines are shown. 
 The parameters are set to $D=D_\theta=\eta=s=1$ and $\ell\to 0$. }
 \label{fig:fig3}  
 \end{figure}

\begin{figure}
 \includegraphics[width=\columnwidth]{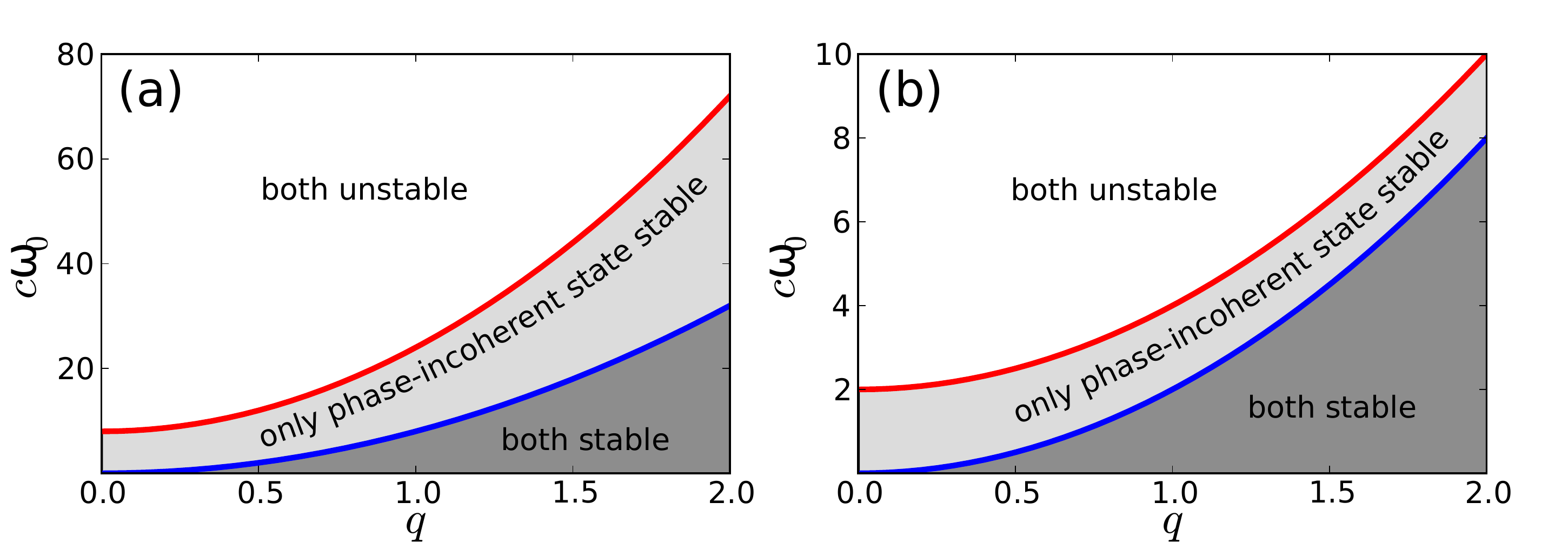} 
 \caption{Phase diagram for (a) shakers and (b) nematic rotors, showing regions of stability of the totally synchronized state (dark grey) and totally phase incoherent state (light grey),  along the direction of fastest growth of the instability, i.e. $q_\perp=q_{||}$ for (a) and $q=q_{||}$ for (b). In the white region time-persistent patterned solutions are to be expected.
In both cases we chose 
$s(\theta)=\cos(\theta)$ and $X(\theta)=-2\sin(\theta)$  
The other parameters are set to $D=D_\theta=\eta=1$ and $\ell\to 0$.}
 \label{fig:fig4}  
 \end{figure}

We present a generic hydrodynamic theory of active particles 
in a fluid medium, accounting for their phases and thus extending earlier
theories for phase-incoherent active fluids.
To this end, we reformulate low-Reynolds-number active-matter hydrodynamics in
terms of phase-dependent force and torque dipoles and obtain generic expressions
for their phase dynamics using the tools of nonequilibrium thermodynamics
\cite{degr}. Our work generalizes the theory of hydrodynamic synchronization
\cite{uchi1011} to oriented and spatially extended systems.
Within this framework we study the stability of the phase incoherent state of a
suspension of active particles and identify conditions under which the particles
synchronize. We consider: (i) active shakers which produce periodic
contributions to the symmetric part of the active stress; (ii) active polar
rotors which produce periodic contributions to the antisymmetric part of the
active stress; and (iii) active nematic rotors which produce periodic
contributions to the active angular momentum flux (Fig.~\ref{fig:fig2}).
We find that all three systems can synchronize by hydrodynamic
interactions (Fig.~\ref{fig:fig3}).
We find, however, that the fully synchronized homogeneous steady state is
generically unstable at zero wavenumber and conclude that the broken
time-translation symmetry gives rise to a class of flow instabilities absent in
the phase-incoherent state of active matter (Fig.~\ref{fig:fig4}).
%
We expect that the phenomena we predict will be seen in experiments on
active-particle fluids at high concentrations, such as realizations of the
oscillating active filaments of \cite{lask13}, and optically driven arrays of
colloidal oscillators or rotors \cite{kota2010,leon12}.

Consider a suspension of active particles which carry a phase variable. 
%
In the absence of external forces and torques, the linear and angular
momentum balance equations of a low Reynolds number fluid are given by
\bea
\partial_\beta(\tilde\sigma_{\alpha\beta}+\sigma^a_{\alpha\beta}
+\sigma^e_{\alpha\beta})&=&0\\
\frac{1}{2}\partial_\gamma(M_{\alpha\beta\gamma}+M^e_{\alpha\beta\gamma})&=&\sigma_{\alpha\beta}^a\quad,
\eea
where $\tilde\sigma_{\alpha\beta}$ and $\sigma^a_{\alpha\beta}$ are the symmetric and antisymmetric parts 
of the deviatoric stress, respectively, and $M_{\alpha\beta\gamma}$ is the deviatoric angular momentum flux. 
Furthermore, $\sigma^e_{\alpha\beta}$ and $M^e_{\alpha\beta\gamma}$ are the hydrostatic contributions to the stress and angular momentum flux, respectively, see \cite{dege95,fuer12tc,fuer12ac}.  

The deviatoric stress and angular momentum flux consist of contributions stemming from the fluid (labelled $f$) and the particles (labelled $act$):
$\tilde\sigma_{\alpha\beta}=\tilde\sigma^f_{\alpha\beta}+\tilde\sigma^{act}_{\alpha\beta}$, $\sigma^a_{\alpha\beta}=\sigma^{a,f}_{\alpha\beta}+\sigma^{a,act}_{\alpha\beta}$ and
$M_{\alpha\beta\gamma}=M^f_{\alpha\beta\gamma}+M^{act}_{\alpha\beta\gamma}$.
%
In the following we consider an incompressible fluid which obeys the constitutive equations,
$\tilde\sigma^f_{\alpha\beta}=2\eta u_{\alpha\beta}$,
$\sigma^{a,f}_{\alpha\beta}=2\eta^\prime(\Omega_{\alpha\beta}-\omega_{\alpha\beta})$ and 
$M^f_{\alpha\beta\gamma}=2\kappa\partial_\gamma\Omega_{\alpha\beta}$,
where $\eta$, $\eta^\prime$ are viscosities and $\kappa$ is the internal rotational friction of the fluid. Here we introduced the strain rate $u_{\alpha\beta}=(\partial_\alpha v_\beta+\partial_\beta v_\alpha)/2$, the vorticity $\omega_{\alpha\beta}=(\partial_\alpha v_\beta-\partial_\beta v_\alpha)/2$, where $\mathbf{v}$ is the center of mass velocity of the fluid, and $\Omega_{\alpha\beta}$ is the intrinsic 'spin' rotation rate  of the fluid \cite{star05,fuer12ac,fuer13}. 

The active symmetric stress generated by a collection of $N$ particles is 
\be 
\tilde\sigma_{\alpha\beta}^{act}=\sum\limits_{i=0}^N D^{(i)}_{\alpha\beta} \delta(\mathbf{r}-\mathbf{r}^{(i)})\quad, 
\ee
where $\mathbf{r}^{(i)}$ is the position of the $i$-th particle and  $D_{\alpha\beta}^{(i)}$ is its intrinsic force dipole \cite{bren77,pedl92}, generated by an underlying periodic process. 
We define the associated phase variable $\theta_0^{(i)}$ by 
\be
D_{\alpha\beta}^{(i)}\equiv s_0(\theta^{(i)}_0)\dot\theta^{(i)}_0 Q_{\alpha\beta}^{(i)}\quad,\label{eq:fdipole} 
\ee
where $s_0(\theta_0^{(i)})$ is the phase dependent action of the swimming stroke.
The axis of the dipole is given by the 
nematic tensor $Q_{\alpha\beta}^{(i)}=p^{(i)}_\alpha
p^{(i)}_\beta-\delta_{\alpha\beta}/d$, where the unit vector $\mathbf{p}^{(i)}$
denotes the particle orientation and $d$ is the number of spatial dimensions.
See Fig. \ref{fig:fig2}a. 

Similarly the antisymmetric active stress and the active angular momentum flux
generated by a collection of $N$ particles are given by 
\bea 
\sigma_{\alpha\beta}^{a,act}&=&\sum\limits_{i=0}^N \tau^{(i)}_{\alpha\beta} \delta(\mathbf{r}-\mathbf{r}^{(i)})\quad,\\ 
M_{\alpha\beta\gamma}^{act}&=&\sum\limits_{i=0}^N T^{(i)}_{\alpha\beta\gamma}\delta(\mathbf{r}-\mathbf{r}^{(i)})\quad,
\eea
where $\tau^{(i)}_{\alpha\beta}$ and $T^{(i)}_{\alpha\beta\gamma}$ are the internal torque and
the torque dipole generated by the $i$-th particle, respectively \cite{fuer12ac,fuer13}. In parallel to Eq.~\ref{eq:fdipole} we define the phases $\theta_1^{(i)}$ and $\theta_2^{(i)}$ of the periodic processes generating active internal torques and angular momentum fluxes, respectively: 
\bea
\tau^{(i)}_{\alpha\beta}&\equiv& s_1(\theta^{(i)}_1)\dot\theta^{(i)}_1 \epsilon_{\alpha\beta\gamma}p_\gamma^{(i)}\\
T^{(i)}_{\alpha\beta\gamma}&\equiv& s_2(\theta^{(i)}_2)\dot\theta^{(i)}_2 m_{\alpha\beta\gamma}^{(i)}
\eea
where $m^{(i)}_{\alpha\beta\gamma}=\epsilon_{\alpha\beta\nu}p^{(i)}_\nu p^{(i)}_\gamma$,  
with corresponding phase dependences carried by functions $s_1$ and $s_2$. See
Fig. \ref{fig:fig2}b, c. 

We obtain dynamical equations for the particle phase variables 
following the logic outlined in \cite{mart72,krus0405,fuer12ac}. The power
dissipated by a
single active particle
embedded in a fluid at temperature $T$ is 
\bea   \label{eq:entrate}
T\dot\Phi&=&s_0\dot\theta_0^{(i)} Q^{(i)}_{\alpha\beta}u_{\alpha\beta}
+s_1\dot\theta^{(i)}_1 m^{(i)}_{\alpha\beta\gamma}\partial_\gamma\Omega_{\alpha\beta}\nn\\
&+&s_2\dot\theta^{(i)}_2\epsilon_{\alpha\beta\gamma}p^{(i)}_\gamma(\Omega_{\alpha\beta}
-\omega_{\alpha\beta})
+r^{(i)}\Delta\mu\quad,   
\eea
where $r^{(i)}$ denotes the rate at which the $i$-th particle consumes fuel
from its energy reservoir and $\Delta\mu$ is the amount of free energy carried
by one unit of fuel. Note that in cases where the $\theta^i_A$
originate from a common periodic process, additional passive couplings
amongst them need to be considered. We ignore these, and the corresponding
contributions to (\ref{eq:entrate}) and (\ref{eq:constitutive1}) below, 
for simplicity. We write dynamic equations for the phase variables by expanding
the thermodynamic fluxes $s_0\dot\theta_0$, $s_1\dot\theta_1$
and $s_2\dot\theta_2$ in terms of the thermodynamic forces
$Q_{\alpha\beta}u_{\alpha\beta}$, $\epsilon_{\alpha\beta\gamma}p_\gamma (\Omega_{\alpha\beta}
-\omega_{\alpha\beta})$ and
$m_{\alpha\beta\gamma}\partial_\gamma\Omega_{\alpha\beta}$. We work to linear
order in the forces, yielding
\bea
s_A\dot\theta_A^{(i)}&=&\zeta_A\Delta\mu+X_{A0} u_{\alpha\beta}
Q^{(i)}_{\alpha\beta}
+X_{A1}\epsilon_{\alpha\beta\gamma}p^{(i)}_\gamma
(\Omega_{\alpha\beta}-\omega_{\alpha\beta})\nn\\
&+&X_{A2}m^{(i)}_{\alpha\beta\gamma}\partial_\gamma\Omega_{\alpha\beta}
+s_A\xi_A(t)
\label{eq:constitutive1}
\eea
where $A$ ranges over $0, 1, 2$ and Onsager's reciprocity principle~\cite{footnote_ons} constrains $X_{AB}$ to be
symmetric. 
In general the phenomenological $X_{AB}$ and $\zeta_A$ are periodic functions of the
phases $\theta_A$.
In Eqns.~
(\ref{eq:constitutive1}),
$\xi_A(t)$ are phenomenological Gaussian white noise
sources.

Finally, to fully specify the system, we provide dynamic equations
 for the particle positions and directors 
\bea
\partial_t\mathbf{r}^{(i)}&=&\mathbf{v}+\mathbf{\xi}_r(t)
\label{eq:rdyn}\\
\partial_t\mathbf{p}^{(i)}&=&\mathbf{h}^{(i)}+
\mathbf{\Omega}\cdot \mathbf{p}^{(i)} +
\nu_1 \bsf{u}\cdot \mathbf{p}^{(i)}         +
\nu_3   \mathbf{p}^{(i)}  \mathbf{p}^{(i)} \cdot \bsf{u}\cdot \mathbf{p}^{(i)}
\nn\\
&+& \nu_4   \mathbf{p}^{(i)}  \mathbf{p}^{(i)}: \bsf{u} \mathbf{p}^{(i)}
+\nu_2(\mathbf{\Omega} -\mathbf{\omega})\cdot\mathbf{p}^{(i)}
+\mathbf{\xi}_\Omega\quad,\label{eq:pdyn}
\eea 
where the coefficients $\nu_1$ to $\nu_4$ describe the particles' tendency to
align with
shear and rotational flow, respectively, and $\mathbf{h}^{(i)}$ is the orientational molecular field aligning
the $i$-th particle with its neighbours \cite{ezhi13}. The functions $\zeta_r(t)$ and $\zeta_\Omega(t)$ are phenomenological white noise sources. 

%
We next seek to understand if active particles 
can synchronize their phases by hydrodynamic interactions.
For simplicity, we restrict our analysis to the case where global
orientational order is not perturbed,
i.e. $p^{(i)}_\alpha=p_\alpha$, $Q^{(i)}_{\alpha\beta}=Q_{\alpha\beta}$ and
$m^{(i)}_{\alpha\beta\gamma}=m_{\alpha\beta\gamma}$ are
imposed and held constant. Such a description will apply on scales smaller than the length $\sqrt{K/\zeta_0 \Delta \mu}$ beyond which the active Freedericksz instability \cite{simh02,voit05,rama07} sets in, where $K$ is a typical Frank \cite{dege95} elastic constant. 

We consider three different cases: 
(i) a suspension of shakers which generate a periodic force dipole only, i.e 
$\zeta_0= \zeta(\theta_0)$, $X_{00} = X_{00}(\theta_0)$ and all other $X$ 
and $\zeta$ are zero, (Fig.~\ref{fig:fig2} (i));
(ii) a suspension of polar rotors which produce an intrinsic torque only, 
i.e. $\zeta_1= \zeta(\theta_1)$, $X_{11} = X_{11}(\theta_1)$ and all other $X$ 
and $\zeta$ are zero, (Fig.~\ref{fig:fig2} (ii)); 
and (iii) a suspension of nematic rotors which produce a torque dipole only,
i.e. $\zeta_2= \zeta(\theta_2)$ and $X_{22} = X_{22}(\theta_2)$ and all other
$X$ 
and $\zeta$ are zero, (Fig.~\ref{fig:fig2} (iii)).   

We start by investigating case (i). 
The phase of the $i$-th shaker, which we shall simply call $\theta^i$, 
obeys
\be
\dot\theta^{(i)}=\omega(\theta^{(i)})+X(\theta^{(i)}) u_{\alpha\beta} Q_{\alpha\beta}+\xi_0(t)\quad,
\ee
where $X=X_{00}/s_0$ and $\omega=\zeta_0\Delta\mu/s_0$, and the active stress
$\tilde\sigma_{\alpha\beta}^{act,(i)}=s_0(\theta^{(i)})\dot\theta^{(i)}
Q_{\alpha\beta}$.

To describe synchronization in a continuum theory we apply the treatment of
\cite{bert0613} to the distribution in position and phase-angle space,
through the Fourier components $Z_n=\sum_i\exp(\i n
\theta^{(i)})\delta(\mathbf{r}-\mathbf{r}^{(i)})$. Here $Z_0(\mathbf{r})$ is
the shaker concentration and $Z_1(\mathbf{r})$ is the complex Kuramoto order
parameter of the suspension. Note that we use $\i$ to denote $\sqrt{-1}$ to
avoid confusion with the index $i$. The dynamic equations for $Z_n$, obtained
from (\ref{eq:constitutive1}) and (\ref{eq:rdyn}), are
\be
(\partial_t+v_\alpha\partial_\alpha -D\Delta+n^2 D_\theta)Z_n=\i n (\omega_m +X_m u_{\alpha\beta}Q_{\alpha\beta}) Z_{n+m}\quad,
\label{eq:zdyn}
\ee
where $D=<\xi_r^2>$ and $D_\theta=<\xi_0^2>$ are translational and
phase-rotational diffusivities, which encode the dephasing effects of the
noise terms in (\ref{eq:rdyn}) and (\ref{eq:constitutive1}). We have expressed
the phase-dependent phenomenological coefficients $\omega$, $X$ and $s$ in terms
of their Fourier coefficients: $\omega(\theta^{(i)})=\sum_n \omega_n \exp(\i n
\theta^{(i)})$ and similarly for $s$ and $X$. The active stress produced by a
suspension of shakers can then be rewritten as 
\bea
\tilde\sigma^{act}_{\alpha\beta}&=&Q_{\alpha\beta}\sum_{n,m}s_n(\omega_m+X_m u_{\alpha\beta}Q_{\alpha\beta}) Z_{n+m}\nn\\&-&\i Q_{\alpha\beta}\sum_n n s_n D_\theta Z_n\quad.
\eea

In this framework, we analyse the linear stability of a
fully phase-disordered suspension around the quiescent homogeneous
steady state $\mathbf{v}=0$, $\Omega_{\alpha\beta}=0$, $Z_0=c$, and all other
$Z_n=0$. We truncate Equations~\ref{eq:zdyn}, or analogous equations for polar and nematic rotors,  for $|n|\ge 2$ in a manner
analogous to \cite{bert0613}. We find that the fully desynchronized state
of the suspension
has a linear instability. If we impose for simplicity $s(\theta)=S \cos(\theta)$ and $\omega(\theta)=\omega_0$,
the stability criterion reads
\be
c \omega_0 S A^{(J)}(\mathbf{q} )\mathrm{Im}(-X_{1})\ge 2 Dq^2+D_\theta\left[2+c S A^{(J)}(\mathbf{q})\mathrm{Re}(X_1)\right]
\ee
where $X_1$ is the first Fourier component of $X(\theta^{(i)})$, and the
superscript $J = i,\, ii,\, iii$ with
\be
A^{(i)}(\mathbf{q})=\frac{ q^2_{||}q^2_\perp}
{ q^4\left(\eta+\eta^\prime
\frac{q^2\ell^2}{q^2\ell^2+1}
\right)- 2 c S  q^2_{||}q^2_\perp\mathrm{Re}(X_1)} \quad,
\label{eq:stability_asynci}
\ee
for shakers, where $q^2=q_{||}^2+q_\perp^2$ with $q_\perp$ and $q_{||}$ denoting the $\mathbf{q}$ directions perpendicular and parallel to the polarity vector $\mathbf{p}$, respectively;  
\be 
A^{(ii)}(\mathbf{q})=\frac{q_{\perp}^2\frac{q^2\ell^2}{q^2\ell^2+1}}
{q^2\left(\eta+\eta^\prime\frac{q^2\ell^2}{q^2\ell^2+1}\right)-2 c S q_\perp^2\frac{q^2\ell^2}{q^2\ell^2+1}\mathrm{Re}(X_1)}\quad.
\label{eq:stability_asyncii}
\ee
for polar rotors, and 
\be
A^{(iii)}(\mathbf{q})=\frac{q_{||}^2}{q^2(q^2\ell^2+1)\left(\eta+\eta^\prime\frac{q^2\ell^2}{q^2\ell^2+1}\right)-2 c S q_{||}^2  \mathrm{Re}(X_1)} \quad,
\label{eq:stability_asynciii}
\ee
for nematic rotors, respectively.

Thus, at large enough driving frequency $\omega_0$ 
and particle density $c$, the fully asynchronous states of suspensions of all
three types of particles is linearly unstable, provided $\mathrm{Im}(-X_1) > 0$,
for the choice $s(\theta) = S \cos \theta$, the case treated here. In general
the synchronization criterion involves phase-dependent stresses and
angular-momentum fluxes $s_A$, and the phase relation between the forcing
$\omega$ and the feedback $X$ from the fluid. The growth rate of the instability
is nonzero for $q \ell \to 0$ for (i) and (iii) and $\mathcal{O}(q^2\ell^2)$ for
case (ii), where $\ell=\sqrt{\kappa/(2\eta^\prime)}$ is  an inherent length
scale which is in general microscopic \cite{star05,fuer12ac}.
Fig.~\ref{fig:fig3} shows the intricate dependence of the
synchronization-threshold value of $c\omega_0$ on the direction of the
perturbation wave vector.

We next investigate the linear stability of the homogeneous phase-coherent
suspension. We define the coarse-grained phase $\Theta(\mathbf{r})=(1/N)
\sum_i^N \theta^{(i)}\delta(\mathbf{r}-\mathbf{r}^{(i)})$ which, in the case
(i) of active shakers, obeys 
\be 
\partial_t\Theta+v_\alpha\partial_\alpha\Theta=D\Delta\Theta+\omega(\Theta)+ X(\Theta)u_{\alpha\beta} Q_{\alpha\beta}\quad,
\ee
with active stress
\be
\tilde\sigma^{act}_{\alpha\beta}=c s(\Theta)Q_{\alpha\beta} \dot\Theta,
\ee
where $c$ is the particle concentration field.

We perturb the homogeneously synchronized quiescent state
$\Omega_{\alpha\beta}=0$ and $\mathbf{v}=0$, $\Theta=\Theta_0(t)$, $c=$
constant, and ask whether the flows set up by the resulting stresses drive the
system further from the unperturbed state. As the reference state is
time-periodic, our result is expressed in cycle-averaged terms. We find that the
system is linearly unstable if
\be
A^{(i)}(\mathbf{q})c \overline{X [s\omega]^\prime}>Dq^2\quad,
\label{eq:stability_synchronized}\ee
where the bar denotes averaging over time 
and the prime denotes a derivative with respect to the argument $\Theta$.
Thus, both asynchronous states and global synchrony are unstable at strong
enough activity, suggesting that the likely long-time behavior consists of
complex spatially patterned states. Stability criteria completely analogous
to (\ref{eq:stability_synchronized}) can also be derived for (ii) and (iii).
In Fig.~\ref{fig:fig4} we plot the phase diagram for solutions for shakers and
nematic rotors. Independent of the precise numerical values of parameters, the
topology of the phase diagram is a robust testable result. Further, we predict
that the zone where only the phase-incoherent state is stable shrinks with
decreasing $D_{\theta}$. Interestingly,
the instability condition Eq.~{(\ref{eq:stability_synchronized})} is always 
satisfied when (\ref{eq:stability_asynci}) holds. Thus a spatially unbounded 
oriented active fluid that synchronizes will do so in a spatially inhomogeneous
manner. 

%


In summary, we have developed a theory for bulk active fluids that accounts for the
underlying periodicity of active processes and displays an instability
towards phase synchronization via a purely hydrodynamic mechanism. We
have mapped out the wave vector dependence of the onset of this synchronization
instability. (Fig.~\ref{fig:fig3}). 
For shakers and nematic rotors, the growth rate of the instability
is non-vanishing for wavenumber $q \to 0$, presumably because of
long-range hydrodynamic interactions carried by the velocity field. In contrast
the growth rate of the synchronization instability of polar rotors vanishes as
$q \to 0$, suggesting that the more rapidly decaying spin rotation rate is
responsible. We then showed that the globally synchronized
quiescent state is also generically unstable (Fig.\ref{fig:fig4}).
Thus spontaneously broken time-translation invariance ultimately leads to
spatial pattern formation and spontaneous flow in bulk active fluids, via
gradients in the Kuramoto phase, even while global nematic order is maintained.
The topology of the resulting phase diagram (Fig.\ref{fig:fig4}) is a  robust
experimentally testable result. We also find that the width of the region in
which the fully phase-incoherent state alone is stable decreases as the
phase-rotational diffusivity decreases. These ideas should be testable in
a variety of systems \cite{kota2010,leon12,lask13}.

In order to focus on general principles we have worked with spatially unbounded
ordered systems, but generalization to thin-film or wall-bounded geometries
\cite{brum12,leon12,vilf06,uchi1011,woll11} is straightforward. Moreover, a
complete treatment must allow distortions of the state of orientational order.
Such a study would bring out the interplay between the spontaneous
flow instabilities \cite{marc13} of asynchronous active liquid crystals and the
new mechanisms outlined in this work. Finally, large-scale numerical studies
will ultimately be required in order to go beyond the linear stability analysis
presented here.  
          
SF was supported by DFG grant FU-961/1-1, and SR acknowledges a J C Bose Fellowship. 
We thank M. Leoni and T. B. Liverpool for sharing a manuscript on a related topic and agreeing to simultaneous submission.

\end{document}